\documentclass[aps,prl,groupedaddress,showpacs, twocolumn, raggedbottom,
nobalancelastpage, superscriptaddress,floatfix]{revtex4}

\usepackage{amsmath}
\usepackage{amssymb}
\usepackage{graphicx}

\newcommand{\up}{U^{\prime}}
\newcommand{\ut}{\tilde{U}}
\newcommand{\upt}{\tilde{U}^{\prime}}
\newcommand{\Tkt}{T_{K}^{SU(2)}}
\newcommand{\Tkf}{T_{K}^{SU(4)}}
\newcommand{\Tk}{T_{K}}
\newcommand{\cm}{\chi^{-}_{c}}

\newcommand{\cs}{\chi_{s}}

\begin{document}
\title{Quantum phase transition in capacitively coupled double quantum dots}
\author{Martin R. Galpin}
\author{David E. Logan}
\affiliation{Oxford University, Physical and Theoretical Chemistry
Laboratory, South Parks Road, Oxford OX1 3QZ, UK.}
\author{H. R. Krishnamurthy}
\affiliation{Department of Physics, IISc, Bangalore 560 012, India.}

\date{\today}

\begin{abstract}
We investigate two equivalent, capacitively coupled semiconducting quantum
dots, each coupled to its own lead, in a regime where there are two electrons
on the double dot. With increasing interdot coupling a rich range of behavior is uncovered: first a crossover from spin- to charge-Kondo physics, via an intermediate $SU(4)$ state with entangled spin and charge degrees of freedom; followed by a quantum phase transition of Kosterlitz-Thouless type to a non-Fermi liquid `charge-ordered' phase with finite residual entropy and anomalous transport properties. Physical arguments and numerical renormalization group methods are employed to obtain a detailed understanding of the problem.
\end{abstract}
\maketitle

\emph{Introduction}. -- 
Semiconducting quantum dots provide \cite{kou} a beautifully direct, tunable mesoscopic realization of a classic paradigm in many-body theory: the spin-Kondo effect \cite{ach}, wherein a single spin in an odd-electron dot is quenched by coupling to the conduction electrons of a metallic lead.
Recent advances in nanofabrication techniques now also permit the controlled
construction of \emph{coupled} quantum dot systems, the simplest being 
double dot (DD) devices. Central to the design of circuits for logic and quantum information processing, and widely studied both theoretically \cite{golden,poh,boe,borda,lopez,andrei,garst} and experimentally \cite{waugh,molen,blick,holl,wilh,chan}, spin \emph{and} orbital degrees of freedom are now relevant, leading to the possibility of creating novel correlated electron states. Recently for example a symmetrical, capacitively coupled semiconducting DD has been studied \cite{borda} in a regime with a single electron ($n=1$) on the DD, and the lowest energy states 
$(n_{L},n_{R}) = (1,0)$ and $(0,1)$ near degenerate. The low-energy physics, which determines the conductance at small bias, was shown \cite{borda} to be governed by a fixed point with $SU(4)$ symmetry, leading to an unusual strongly correlated Fermi liquid state where the spin and orbital degrees of freedom are 
entangled.

In this paper we study a capacitively coupled, symmetrical semiconducting
DD system, but now in a regime with two electrons on the DD such that $(1,1)$,
$(2,0)$ and $(0,2)$ are the relevant low energy states. As shown below, the associated physics is both rich and qualitatively distinct from the $n=1$ sector: on increasing the ratio $U^{\prime}/U$ of inter- and intra-dot coupling strengths, we find that
the system first evolves continuously from an $SU(2) \times SU(2)$ spin-Kondo state
where the dot spins are in essence separately quenched,
to an $SU(4)$ Kondo state with entangled charge and spin degrees of freedom when
$U^{\prime}/U=1$. Thereafter, for a tiny increase in $U^{\prime}/U$, 
there is then a smooth crossover to a novel charge-Kondo state;
followed by suppression of charge-pseudospin tunneling, manifest in collapse of the associated Kondo scale and a Kosterlitz-Thouless (KT) quantum phase transition to a  doubly degenerate charge-ordered state --- a non-Fermi liquid 
phase with $\ln 2$ entropy 
and anomalous low-energy transport and thermodynamic properties. 
Detailed results for this diverse range of behavior are obtained using the numerical renormalization group (NRG) method \cite{kgw,hrk}, preceded by simple
physical arguments that enable the essential physics to be understood.

\emph{Model and physical picture}. -- 
We consider two equivalent, capacitively coupled semiconducting (single-level) dots,
each coupled to its own lead. The Anderson-type Hamiltonian
is $H = H_{0} + H_{V} +H_{D}$, where $H_{0} = \sum_{i,\mathrm{k},\sigma}\epsilon_{\mathrm{k}}a_{\mathrm{k}i\sigma}^{\dagger}a_{\mathrm{k}i\sigma}^{\phantom\dagger}$ 
refers to the leads ($i =L/R$) and $H_{V}=\sum_{i,\mathrm{k},\sigma}V(a_{\mathrm{k}i\sigma}^{\dagger}c_{i\sigma}^{\phantom\dagger} + h.c.)$ to the lead-dot couplings. 
$H_{D}$ describes the isolated dots,
\begin{equation} 
H_{D}=\sum_{i=L,R}(\epsilon\hat{n}_{i} + U\hat{n}_{i\uparrow}\hat{n}_{i\downarrow})+
\up\hat{n}_{L}\hat{n}_{R} 
\end{equation}
with $\hat{n}_{i} = \sum_{\sigma}\hat{n}_{i \sigma} =
\sum_{\sigma}c_{i\sigma}^{\dagger}c_{i\sigma}^{\phantom\dagger}$.
$U$ denotes the intradot Coulomb interaction, $\up$ the interdot (capacitive) coupling. 
In the isolated DD, increasing $|\epsilon| = -\epsilon$ (via suitable gate voltages) generates the usual Coulomb blockade staircase. For $0 < |\epsilon| 
<\mathrm{min}(U,\up)$ the ground state occupancy is $n=1$, with degenerate configurations $(n_{L},n_{R}) =(1,0)/(0,1)$ and a
$\ln4$ residual entropy $(k_{B} \equiv 1)$ \cite{poh,boe,borda}. Coupling to the leads
quenches this entropy, and the strongly correlated effective low-energy model is $SU(4)$ Kondo \cite{boe,borda}. 
The underlying physics here is rich, including spin-filtering arising from the
continuous crossover to the $SU(2)$ orbital Kondo effect in a strong magnetic field
\cite{borda}. But no quantum phase transition occurs in this $n=1$ sector.

  We consider by contrast the $n=2$ domain of the Coulomb 
staircase, arising for min$(U,\up) < |\epsilon| < \up +$ max$(U,\up)$.
Two sets of configurations then dominate, according to whether $\up \lessgtr U$:
the 4-fold spin-degenerate states $(n_{L},n_{R}) =(1,1)$, 
and the degenerate pair $(2,0)/(0,2)$, with DD energy difference 
$E(2,0)-E(1,1) = U-\up$. The DD ground state is thus $(1,1)$ for $\up<U$, and 
$(2,0)/(0,2)$ 
for $\up >U$; all six states are degenerate at $\up =U$ where the model has $SU(4)$ symmetry. We first give physical arguments for the evolution of the coupled DD-lead system with increasing $\up$; focusing on the strongly correlated regime of
$U/\Gamma \gg 1$ ($\Gamma = \pi V^{2}\rho$ with $\rho$ the lead DoS).
Here an effective low-energy Hamiltonian may be obtained from second order perturbation theory (PT) in the lead-dot tunneling $V \equiv V_{L}
=V_{R}$ (with $V_{L/R}$ denoting coupling to the $L/R$ lead)~\cite{fnn}.

 For $\up =0$ the dots are fully decoupled. Only $(1,1)$ states are relevant.  
The effective model is obviously two uncoupled spin-$\tfrac{1}{2}$ Kondo models.
The spin entropy is quenched at the normal
Kondo scale $\Tkt$, leading to a local singlet ground state (`$SU(2)\times SU(2)$ 
spin-Kondo').
For $\up \gg U$ by contrast the $(2,0)/(0,2)$ DD states dominate, and
as shown below the ground state is a \emph{doubly degenerate} `charge ordered' (CO) state with $\ln2$ entropy. Continuity then implies a quantum phase transition at some critical $\up_{c}$. As discussed below, for $\up =U$ the effective model is $SU(4)$ Kondo (in the $n=2$ sector) \cite{fnn}, with entangled spin/charge degrees of freedom
but a singlet ground state with a larger Kondo scale $\Tkf$; and is connected
continuously to the $SU(2)\times SU(2)$ spin-Kondo state arising as
$\up \rightarrow 0$. We thus expect $\up_{c}>U$.

Hence consider increasing $\up$ above $U$.
Since the configurations $(1,1),(2,0),(0,2)$ are
degenerate for $\up =U$ this full $n=2$ manifold must be retained for $\up \simeq U$.
Virtual excitations to excited states are eliminated via PT, and divide
in two classes \cite{fnn}: (a) $V_{L}^{2}$ or $V_{R}^{2}$ processes, involving tunneling to one
lead alone. \emph{Any} configuration connects to itself via such, \emph{eg}
$(1,1) \leftrightarrow (1,1)$ under $V_{R}^{2}$ via excited states $(1,0)$ or
$(1,2)$. (b) $V_{L}V_{R}$ processes. These necessarily connect different manifold configurations; the full set is clearly 
$(2,0) \leftrightarrow (1,1)$
and $(0,2) \leftrightarrow (1,1)$, there being no direct 
coupling between $(2,0)$ and $(0,2)$. As $\up$ increases above $U$
charge and spin states begin to separate: the degenerate charge pair $(2,0)/(0,2)$,
components of an effective charge pseudospin, lie lower in energy by $\up -U$
than the $(1,1)$ spin states. For sufficiently small $\up -U>0$, tunneling
between the $(2,0)/(0,2)$ states can however still arise, and quench the charge pseudospin (and 
hence entropy), producing thereby a non-degenerate \emph{charge-Kondo} state. 
But, as above, this tunneling is not direct, being mediated by the 
higher energy $(1,1)$ states.
We thus expect the associated Kondo scale to be diminished compared to
$\Tkf$ (the `stabilization' due to dot-lead coupling at the $SU(4)$ point)
and to decrease as $\up$ increases; and moreover that the quenching will cease
to be viable when the relative energy $\up -U$ of the $(1,1)$ states exceeds
roughly $\Tkf$, leading to a quantum phase transition to the degenerate CO
phase when $\up_{c} -U \approx \Tkf$.
Since the latter is exponentially small for strong correlations, this implies
a critical $\up_{c}$ exponentially close to $U$ (as confirmed by NRG below).

These arguments extend readily to $\up <U$, but now the circumstances differ.
Spin/charge degrees of freedom do 
separate on decreasing $\up$ from $U$, $(1,1)$ states now lying lower by
$U-\up$ than $(2,0)/(0,2)$. But since the $(1,1)$ spin-states connect \emph{directly}
to themselves under $V_{R}^{2}$ or $V_{L}^{2}$ (as above), quenching of the spin
entropy is not inhibited and no transition occurs.
Instead a continuous crossover from $SU(4)$ Kondo to separable $SU(2)\times SU(2)$ 
spin-Kondo is expected, for $U -\up \approx \Tkf$.

\emph{Results}. --
  The physical picture is thus clear, and the transition to the degenerate CO phase occurs in the vicinity of the $SU(4)$ point $\up =U$ \cite{fnn}.
We now present NRG results for the DD Anderson model, choosing 
the midpoint of the $n=2$ domain, $|\epsilon| = U/2 +\up$. This case is particle-hole (p-h) symmetric, but representative. The physics is
wholly robust to departure from p-h symmetry. 

The low temperature behavior of the model is governed by two
classes of stable fixed points (FP), corresponding to the two zero
temperature phases. The first, a strong coupling (SC) fixed point, 
describes all the singlet ground states; and is reached (as $T \rightarrow 0$ 
or NRG iteration number $N \rightarrow \infty$) for all $\up <
\up_{c}$. The corresponding FP Hamiltonian is simply a
doubled version ($SU(2) \times SU(2)$) of that well known for the
spin-1/2 Anderson model \cite{hrk}. The dot spins and hence
entropy are thus quenched at $T = 0$, the system being a Fermi
liquid and characterized by a Kondo scale denoted generically as
$\Tk$. The second, reached for all $\up > \up_{c}$, is a line (i.e.\ a
one parameter family) of charge ordered (CO) FP. The
generic FP Hamiltonian corresponds to setting $\Gamma =0$
and $\up = \infty$. The DD and leads are then decoupled, but
the FP has internal structure reflecting broken symmetry, since dot-states
occur in the degenerate pair $(n_{L},n_{R}) = (2,0)/(0,2)$ (whence
$\ln2$ residual entropy). The line of FP is
obtained by supplementing the free lead Hamiltonians  by 
potential scattering {\it correlated to dot occupancy},  of form $H_{K} =
K\sum_{i,\sigma}\sum_{\mathrm{k},\mathrm{k^{\prime}}}
a_{\mathrm{k}i\sigma}^{\dagger}a_{\mathrm{k^{\prime}}i\sigma}^{\phantom\dagger}
(\hat{n}_{i}-1)$ (\emph{cf} \cite{fnn}). The actual value of $K$ is
obtained numerically by matching to NRG energy levels.

  A comparison of NRG energy level flows for large iteration number,
with the characteristic energy level structure for the two FP, enables
the phase diagram to be found; as
shown in Fig.~1(a) 
\begin{figure}
\includegraphics{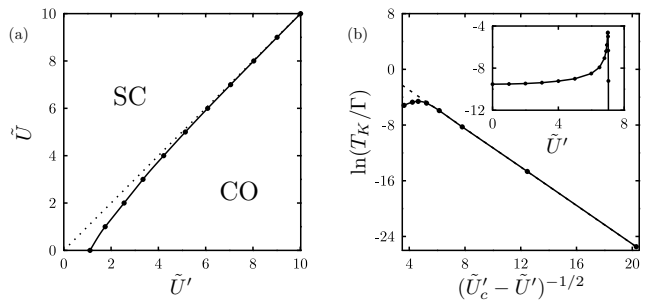}
\caption{(a) Phase diagram in $(\upt,\ut)$-plane;
the $SU(4)$ line $\up =U$ is also shown (dotted). 
(b) For $\ut =7$ in the SC phase, $\ln(\Tk/\Gamma)$ \emph{vs} 
$[\upt_{c}-\upt]^{-1/2}$ close to the transition, showing the exponential vanishing
of $\Tk$. Inset: $\ln(\Tk/\Gamma)$ \emph{vs} $\upt$.}
\end{figure}
in the $(\upt =\up/\pi\Gamma,
\ut = U/\pi\Gamma)$-plane \cite{fna}. The transition is seen to occur for all $\ut \geq 0$ 
on increasing the interdot $\upt$. Consistent with the physical arguments above,
for $\ut \gg 1$ the critical $\up_{c}$ indeed lies exponentially close to the $SU(4)$
line $\up =U$ (specifically we find $(\up_{c}/U-1) \simeq 2\Tkf/\Gamma\ut^{1/2}$). 
Fig.~1(b) (inset) shows the $\upt$-evolution 
of $\Tk$ \cite{fna} in the SC phase, for a typical 
strongly correlated $\ut =7$.
It is seen to depart little from its $\up =0$ value $\Tkt \propto
\Gamma\ut^{1/2}
\exp(-1/\rho J)$ (with $\rho J = 8/\pi^{2}\ut$) \cite{hrk}, indicative
of spin-Kondo physics, until very close to $\up =U$ where it
increases rapidly to $\Tkf \propto \Gamma\ut^{3/4}\exp(-1/2\rho J)$;
\emph{ie} while exponentially small, $\Tkf \propto [\Tkt]^{1/2}$ shows 
a strong relative enhancement at the $SU(4)$ point \cite{boe,borda}. 
However on increasing $\up$ above $U$ and entering the charge-Kondo regime, $\Tk$ is seen to drop rapidly and vanishes as the 
SC$\rightarrow$CO transition is approached. The transition is of KT type, consistent with the line of CO FP for $\up \geq \up_c$, and (from NRG energy level flows) no
evidence for a separate critical FP;
further evidenced by \cite{KT}
the $\upt \rightarrow \upt_{c}-$ behavior $\Tk \propto \exp(-a/[\upt_{c}-\upt]^{1/2})$, demonstrated in Fig.~1(b) for $\ut =7$.
This behavior is generic,
even for $U=0$; here the non-interacting $SU(4)$ ($\up =0$) scale $\Tkf \sim \Gamma$, and $\up_c - U \approx \Tkf$ implies $\upt_c \sim {\cal{O}}(1)$, as indeed 
found (Fig. 1(a)).

In addition to the two stable (low-temperature) FP, three unstable 
FP play an important role at finite-$T$, and are seen clearly in the
impurity entropy $S$  $(\equiv S_{\mathrm{imp}})$: (i) free orbital (FO) \cite{hrk},
corresponding to $\Gamma =0 =U= \up$, with $\ln16$ entropy. This 
is just the high-$T$ limit of $S(T)$, reached in all cases for 
non-universal $T \approx$ max$(U,\up)$.
(ii) $SU(4)$ local moment  (LM$^{SU(4)}$), corresponding to $\Gamma =0$ and
$U =\infty =\up$, with associated entropy $\ln6$; and (iii) $SU(2)\times SU(2)$ 
local moment (LM$^{SU(2)}$), $\Gamma =0 = \up$ and $U=\infty$, with
$\ln2^{2}$ entropy. Fig.~2 shows $S(T)$ \emph{vs} $T/\Gamma$ for $\ut =7$ ($\upt_{c}
\simeq 7.046$). 
\begin{figure}
\includegraphics{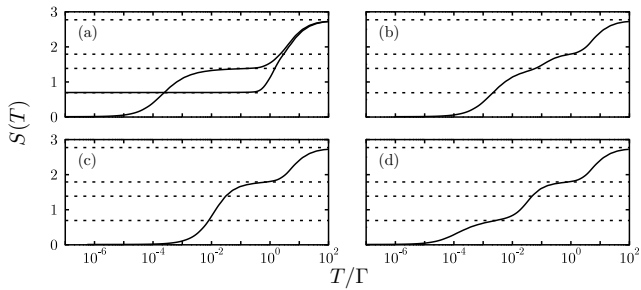}
\caption{$S(T)$ \emph{vs} $T/\Gamma$ for $\ut =7$ ($\upt_{c} \simeq 7.046$).
(a) $\upt =6$ (SC, `uncoupled spin-Kondo') and $8$ (CO); (b) $\upt =6.9$ 
(crossover to $SU(4)$); (c) $\upt =7$ ($SU(4)$); (d) $\upt =7.03$ (charge-Kondo).
Dotted lines show $\ln16$, $\ln6$, $\ln4$ and $\ln2$, associated 
with the FO, LM$^{SU(4)}$, LM$^{SU(2)}$ and CO fixed points respectively.
}
\end{figure}
Fig.~2(a) illustrates the behavior `deep' in the CO and SC phases.
In the former $S(T)$ simply crosses directly from its $\ln2$ (CO) residual value
to $\ln16$ (FO) on the scale $T \sim \up$; while in the latter, consistent with
the effective underlying spin-$1/2$ Kondo physics, there is first a crossover from 
$S(0)=0$ (SC) to $\ln4$ (LM$^{SU(2)}$) for $T \simeq \Tk$ $(\simeq \Tkt)$.
Figure 2(b), for $\upt =6.9 <\ut$, 
illustrates the crossover 
from effective uncoupled $SU(2)$ Kondo to
$SU(4)$. Here $S(T)$ increases in a two-stage fashion, first to 
$\ln4$ (LM$^{SU(2)}$) for $T\sim\Tk$ and then $\ln6$ (LM$^{SU(4)}$) for
$T\sim [U-\up]$ $(= E(2,0)-E(1,1))$. This behavior is naturally absent at the 
$SU(4)$ point, Fig.~2(c): $S(T)$ crosses directly to $\ln6$ for $T\sim \Tkf$.
In the charge-Kondo regime Fig.~2(d) (for $\upt = 7.03$) 
$S(T)$ again shows the two-stage behavior 
typical of a KT transition \cite{garst,hof}, 
but here it first crosses
from $0$ (SC) to $\ln2$ (CO) for $T\sim\Tk$ and then to $\ln6$ (LM$^{SU(4)}$)
for $T \simeq [\up -U] =E(1,1)-E(2,0)$; consistent with the physical discussion
given above.

The physics discussed above naturally shows up also in
various thermodynamic susceptibilities.
For example, as $\upt$ is increased past $\ut$ into the
charge-Kondo regime, we find the `impurity' spin susceptibility $\cs$
decreases monotonically; but its charge pseudospin analogue the 
staggered charge susceptibility $\cm$,
given by $\cm \sim 1/\Tk$, diverges as $\upt \rightarrow
\upt_{c}$ reflecting the collapse of the charge-Kondo state and
the quantum phase transition. For $\upt > \upt_{c}$ the ($T=0$)
$\cm$ remains infinite, symptomatic of the broken symmetry CO
phase, with $\cm(T) \propto 1/T$ as $T\rightarrow 0$. Further
details will be given in subsequent work.

\begin{figure}
\includegraphics{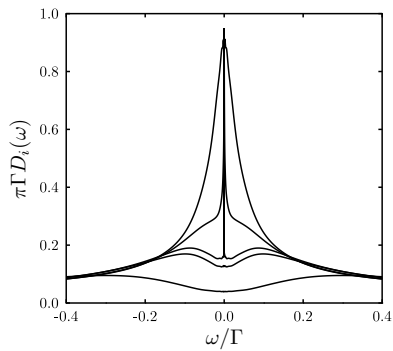}
\caption{
Transmission $\pi\Gamma D_{i}(\omega)$ \emph{vs} $\omega/\Gamma$, for
$\ut =7$ and (top to bottom)
$\upt = 7, 7.03, 7.044$
(SC) and $7.048, 7.1$ (CO).
}
\end{figure}
  Finally and most importantly, the destruction of the Kondo effect as the SC$\rightarrow$CO transition is approached is seen vividly in electronic transport, notably the transmission
coefficient $T_{i}(\omega) = \pi\Gamma D_{i}(\omega)$ with $D_{i}(\omega)$ the $T=0$ local single-particle spectrum ($D_{i}(\omega) = -$Im$G_{i}(\omega)/\pi$ with 
$G_{i}(t) =-i\theta(t) \langle\{c_{i\sigma}^{\phantom\dagger}(t),
c_{i\sigma}^{\dagger} \} \rangle$ the retarded dot Green function).
At the Fermi level in particular ($\omega =0$), 
$T_{i}(0)$ gives
the linear differential conductance across one (either) dot in units of the conductance 
quantum $2e^{2}/h$ \cite{meir}; and at finite, low bias voltage $V$, the equilibrium
$T_{i}(\omega =eV)$ provides an approximation to the conductance \cite{meir}.
The low-energy behavior of $T_{i}(\omega)$ is shown in Fig.~3 for 
a range of $\upt$ spanning the transition. For $\upt =\ut$ a relatively broad Kondo resonance characteristic of the $SU(4)$ point is apparent, with width 
$\propto \Tk = \Tkf$. On increasing $\upt$ into the charge-Kondo regime the Kondo resonance, now residing on top of an incoherent continuum, remains intact with 
$T_{i}(0) =1$ throughout the SC phase reflecting the unitarity limit  
\cite{fnb}. But it narrows progressively as  $\Tk$ diminishes, and as $\upt \rightarrow \upt_{c}-$ the Kondo resonance vanishes `on the spot', such that for $\upt > \upt_{c}$ in the CO phase only the background continuum remains.
The linear conductance in particular thus drops abruptly 
at the transition.
This appears to be a general signature of an underlying KT transition, it being found also for a multi-level small dot close to a singlet-triplet degeneracy point 
\cite{hof}; and in recent work \cite{garst} on two Ising-coupled Kondo impurities, onto which maps the problem of spinless, capacitively coupled metallic islands/large dots close to the degeneracy point between $N$ and $N+1$ electron states \cite{andrei}.
The non-Fermi liquid nature of the CO phase is also seen clearly here, because
$T_{i}(0) = 1/[1+(\Sigma^{\mathrm{I}}(0)/\Gamma)]$ in terms of the imaginary part
of the dot self-energy; whence $T_{i}(0) <1$ in the CO phase implies a non-zero 
$\Sigma^{\mathrm{I}}(\omega =0)$ and thus a non-FL state
(as also manifest \emph{eg} in anomalous exponents for the subleading 
$T$-dependence of thermodynamic properties, although the latter effects are 
quantitatively minor).

  We have considered an equivalent (L/R symmetric) DD system, with specific
results shown for the p-h symmetric case. For a DD device described by the effective model to be realizable, the physics should be suitably robust to
breaking these symmetries. As mentioned previously, that is entirely so
for departure from p-h symmetry.
L/R symmetry can be broken by detuning \emph{eg} the dot levels, $\epsilon_{R/L} =
\epsilon \pm \delta\epsilon$, or coupling to the leads, $\Gamma_{R} \neq \Gamma_{L}$.
In that case it is readily shown that the SC FP remains stable, no new 
corrections to the FP being generated. For the CO FP by contrast, additional
relevant perturbations arise. This FP is thus unstable, and flows in its vicinity
ultimately cross over to the SC FP under renormalization, with the crossover
scale determined by $\delta\epsilon$. We find however that \emph{eg} for small 
but finite $\delta\epsilon \ll \Tkf$, the thermodynamics above
are essentially unaffected for $T \gtrsim \delta\epsilon$;
and that the abrupt drop in the linear conductance at the transition is simply replaced by a continuous but nonetheless sharp crossover over a $\up$ interval 
on the order of $\Tkf$. In that sense the physics described here is thus also 
robust to L/R symmetry breaking.

\emph{Conclusion}. --
Motivated by extensive recent interest in capacitively coupled DD systems
\cite{poh,boe,borda,lopez,golden,andrei,garst,waugh,molen,blick,holl,wilh,chan}, 
we have analyzed a symmetrical, capacitively coupled 
semiconducting DD in the two-electron regime, using the NRG technique.
We have shown that on 
increasing the interdot coupling the system evolves continuously through a 
progression of Fermi liquid states from a purely \emph{spin}-Kondo state, 
via the $SU(4)$ point where charge and spin degrees of freedom are wholly entangled, 
to a \emph{charge}-Kondo state with a quenched charge-pseudospin; before undergoing
a quantum phase transition of the Kosterlitz-Thouless type to a non-Fermi liquid,
doubly degenerate charge-ordered phase. This provides a striking example of the subtle and many-sided interplay between spin and charge degrees of freedom in small quantum dots.


\begin{thebibliography}{99}
\bibitem{fnn}
$H_{\mathrm{eff}}$ 
in the $n=2$ manifold is:
$H_{\mathrm{eff}} - H_{0} = H_{\delta} +H^{\prime}$ where
$H_{\delta} = [U-\up]\sum_{i}\hat{n}_{i\uparrow}
\hat{n}_{i\downarrow}$, vanishing at the $SU(4)$ point; and
$H^{\prime}=
\tfrac{J}{2}\sum_{\mathrm{k},\mathrm{k^{\prime}}}\sum_{i,\sigma}
\sum_{j,\sigma^{\prime}}
a_{\mathrm{k^{\prime}}j\sigma^{\prime}}^{\dagger}a_{\mathrm{k}i\sigma}^{\phantom\dagger}
c_{i\sigma}^{\dagger}c_{j\sigma^{\prime}}^{\phantom\dagger}
$
with an antiferromagnetic exchange $J$.
Terms with $j\neq i$ in 
$H^{\prime}$ arise from $V_{L}V_{R}$
processes, interconnecting $(2,0)/(0,2)$ and $(1,1)$ states. Those
with $j=i$ come from the $V_{R}^{2}$ or $V_{L}^{2}$ processes:
acting on $(1,1)$ states gives rise per se to $SU(2)\times
SU(2)$ spin-$1/2$ Kondo; while the $(2,0)/(0,2)$ states generate
potential scattering of form
$\tfrac{J}{4}\sum_{i,\sigma}\sum_{\mathrm{k},\mathrm{k^{\prime}}}
a_{\mathrm{k}i\sigma}^{\dagger}a_{\mathrm{k^{\prime}}i\sigma}^{\phantom\dagger}
(\hat{n}_{i}-1)$, equal and opposite for $n_{i}=0$ and $2$.
\bibitem{fna} Results are typically shown for an
NRG discretization parameter \cite{hrk} $\Lambda =3$ keeping $\sim 20,000$ states
(excluding spin degeneracy) per NRG step. $\Tk$ is defined in practice 
by $\Tk =1/\gamma$, with $\gamma$ the usual $T=0$ linear specific heat 
coefficient \cite{hrk}.
\bibitem{fnb} The unitarity limit is underestimated by $\sim 10\%$ in NRG due to well known systematic numerical errors.
\end{thebibliography}
\end{document}